\documentstyle[prd,aps,eqsecnum]{revtex}

\input epsf.tex
\tighten
\begin{document}
\draft
\twocolumn[\hsize\textwidth\columnwidth\hsize\csname 
@twocolumnfalse\endcsname

\title{Annual modulation sensitivity in cold dark matter searches}

\author{F. Hasenbalg$^{\rm *}$}


\address{Departamento de F\'{\i}sica, Comisi\'on Nacional de
Energ\'{\i}a At\'omica, Av. del Libertador 8250, 1429 Buenos Aires,
Argentina}

\date{\today}
\maketitle

\begin{abstract}
The sensitivity of experiments looking for annual modulated signals is
discussed and analyzed.  The choice of energy intervals for rate
integration and enhancing the signal-to-noise ratio of the
predicted WIMP signal is addressed.  Special emphasis is placed on
quantifying the minimum required exposure, $MT$, under experimental
conditions.  The difficulty reduces to establish the proper separation
between the rate due to the unmodulated part of the WIMP signal and the
rate of spurious background present in any experiment.  The problem is
solved by placing an upper bound to the unmodulated part of the signal
using the best exclusion plot. We find that the lowest backgrounds
achieved result in exposures in the range $MT \sim 5 - 100 $~kg~yr for
masses $m_{\chi} > 100$~GeV depending on the energy threshold of the
detector.  While our results are valid for Ge and NaI detectors, 
the formulae derived apply to other elements as well.  Prescriptions are
given to estimate exposures for higher background experiments.
\end{abstract}

\vskip2pc]

\section{Introduction}

Weakly interacting massive particles (WIMP's) are most likely the main
constituents of the galactic halo.  Constraints on the strength of the
interaction and mass of the WIMP's have been obtained by several
experiments using germanium \cite{Beck}, silicon \cite{Caldwell2}, and
NaI(Tl) \cite{Bernabei} crystals.  For a coupling constant of the order
of the weak coupling constant, the mass of these WIMP's must be, in any
case, heavier than the 40~GeV lower bound already imposed by the
measurement of the width of the $Z^0$ \cite{Krauss}. The same
measurement has also placed bounds on some supersymmetric candidates.

Drukier et al. \cite{Drukier}, suggested that a distinctive signature
of WIMP's would be the annual modulation of the detection rate. The
modulation originates in the orbital motion of the Earth around the Sun
that produces a variation in the relative velocity between the Earth
and the WIMP's, thus altering the dark matter flux during the year.
Extracting this fluctuation from data that, certainly, contain the
contribution of spurious events is a task that demands care. The
problem to be confronted with is the following:  to extract a feeble
signal from the data, we require a large signal-to-noise ratio so that
the relative statistical fluctuations are small compared with the
sought after signal.  At the same time, efforts to reduce the
background of unwanted (no dark matter) events prevent achieving large
signal-to-noise ratios. There is no compromise solution possible; if a
low counting rate has been attained, a long exposure is required
to detect an even smaller modulation.

The main purpose of this work is to examine this issue {\em quantitatively}
and to establish some guidelines upon which upcoming data, like those
from the Sierra Grande experiment \cite{Digregorio}, can be analyzed.
Although the subject of annual modulation, and its details, have been
discussed by several authors
\cite{Freese,Sadoulet,Gabutti,Cimento,Lewin}, we feel that the
sensitivity of modulation searches, namely the amount of exposure
needed to resolve a given fluctuation embedded in a background, has not
been yet addressed quantitatively in the literature.  A
semi-quantitative approach was advanced in Ref.~\cite{Baudis} where
results for required exposures were presented.  Our procedure to obtain
minimum exposures however, relies on simpler assumptions and give lower
estimates.

In the first section we briefly introduce the formalism needed for
calculating the expected signal-to-noise ratios. For the
sake of simplicity, we illustrate our calculations for the case of a 
germanium detector later on we extend the results to a sodium iodide detector.
We take into account the influence of the crossing
of the spectra due to their annual fluctuation and choose energy
intervals of integration that maximize the expected signal-to-noise
ratios of the modulated rate. The next section analyzes quite generally the
sensitivity of this type of experiments, addressing its difficulties
and common misuses.  Estimates of the exposures for the best
experiment are presented together with guidelines for
experiments with higher backgrounds.  Most of our results are not
strongly influenced by astrophysical uncertainties.

\section{Predicted signal}

The expected total rate of events due to the recoil of nuclei elastically
scattered by WIMP's will be the product of the cross
section, the WIMP's flux, and the number of target nuclei in
a detector of atomic mass number $A$.  For an incoming WIMP of mass
$m_{\chi}$ and velocity $v$ the differential counting rate in the
recoil-energy interval $T$, $T+dT$, is given by

\begin{equation}
\frac{dR}{dT}= \frac{N_A}{A} \left( \frac{\rho_{halo}}{m_\chi}
     \right)  \int_{v_{min}}^{v_{max}} f(v,\sigma_v, V_{\cal E}) v
      \frac{d\sigma}{dT}(v) dv ,   \label{rate}
\end{equation}
where $N_A$ is Avogadro's number, $\rho_{halo}/m_\chi$ is the number
density of WIMP's, and $f(v, \sigma_v, V_{\cal E})$ is the
Maxwell-Boltzmann velocity distribution of the WIMP's for an observer
on the Earth~\cite{Freese}. The velocity distribution is a function of
$V_{\cal E}(t)$, the velocity of the Earth with respect to the galactic
rest frame varying annually and $\sigma_{v}$, the
dispersion velocity of the WIMP's in the galactic halo.  The
integration limits of (\ref{rate}) are, $v_{max}$, the maximum velocity
of the WIMP's ($v_{max} = V_{\cal E} + v_{esc}$, where $v_{esc}$ is the
escape velocity from our galaxy), and $v_{min}$, the minimum WIMP
velocity necessary to contribute to a particular energy of the recoil
spectrum.

The differential cross section in Eq. (\ref{rate}) is given, in general, by

\begin{equation}
  \frac{d\sigma}{dT} \;=\; \frac{\sigma_{\rm o}(m_\chi)}{T_{max}}\, F^2(T),
\end{equation}
where $\sigma_{\rm o}(m_\chi)$ is the cross section at zero momentum
transfer (of a heavy Dirac neutrino, for example), $T_{max} = 2 \mu^2
v^2/m_N$ ($\mu$ is the WIMP-nucleus reduced mass and $m_N$ the nucleus mass),
and $F^2(T)$ is a standard Bessel nuclear form
factor~\cite{Lewin} that depends on the momentum transfer and which
takes into account the loss of coherence of the interaction. The
calculated recoil spectrum must also be convoluted with a relative
efficiency function for germanium~\cite{Ahlen} accounting for the
efficiency of the recoiling process in generating an ionization signal.
We refer therefore to the deposited energy $E$, instead of $T$, in the
following formulae. Table \ref{table1} lists the values of
$\rho_{halo}$, $v_{esc}$, $\sigma_{v}$, and $V_{\odot}$, the velocity
of the Sun around the galactic center, used in this work.

\subsection*{Crossing energies}

It is known that, the differential energy spectra of June and December
cross at specific energies that depend, among other factors, on the
WIMP mass \cite{Sadoulet}.  The reason for the crossing of the spectra
is that in June the velocity distribution  increases by
15~km~s$^{-1}$ enabling the WIMP's to deposit a higher fraction of
their kinetic energy.  This crossing can significantly alter the
amplitude of the modulation depending on the energy interval chosen to
integrate the differential rate.

The calculated crossing energies, $E_c$, (dotted and dashed lines) are
shown in figure 1, for the case of a heavy Dirac neutrino, as a
function of the WIMP mass.   

\begin{table}

\caption{List of parameters used.}
\vspace{.3cm}

\begin{tabular}{c}
$\rho_{halo} = 0.3$~GeV~cm$^{-3}$ \\
$V_{\odot} = 230$~km~s$^{-1}$ \\
$v_{esc} = 570$~km~s$^{-1}$ \\
$\sigma_{v} = 270$~km~s$^{-1}$ \\
\end{tabular}
\label{table1}

\end{table}
\noindent
The figure explores also the dependence of
the crossing energy on different halo parameters. The halo density for
example, has no effect on the crossing energies since it affects only the
absolute rates. The crossing is neither affected by the assumed escape
velocity, $v_{esc}$, or the value of $V_{\odot}$ (which in turn modifies
$V_{\cal E}$) but, it is sensitive to the dispersion velocity, $\sigma_v$,
which for WIMP masses higher than 500~GeV changes by 2~keV, when $\sigma_v$
is increased from 270~km~s$^{-1}$ to 300~km~s$^{-1}$. The dispersion
velocity of WIMP's in the galactic halo is a rather uncertain parameter,
ranging from 246~km~s$^{-1}$ to 323~km~s$^{-1}$~\cite{halo}, so here we adopt
a standard value of 270~km~s$^{-1}$.

\vspace{-1.cm}

\begin{figure}
\centering
\epsfxsize=9.0truecm
\epsfysize=6.8truecm
\epsffile{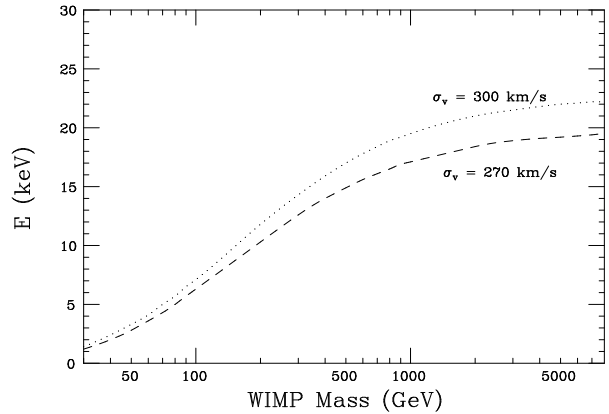}
\caption{Crossing energies in Ge as a function of WIMP mass for
two different dispersion velocities. The dotted line corresponds to a
dispersion velocity $\sigma_v = 300$~km~s$^{-1}$ and the dashed line to
$\sigma_v = 270$~km~s$^{-1}$.\label{fig1}}
\end{figure}

\subsection*{Theoretical signal-to-noise ratio}

To extract a signal from the data it is necessary to establish the
maximum of the theoretical signal-to-noise ratio.  The total signal as
a function of time can be expressed in the form,
\begin{equation}
  S(t) =\int \frac{dR}{dE'}[V_{\cal E}(t)] dE' = 
    S_{\rm o} + S_m \cos(\omega t) + {\cal O}(S_m^2) , \label{st}
\end{equation}
where we define $S_{\rm o}$ as the unmodulated part,
$S_m$ as the amplitude of the mo\-du\-la\-tion, $\omega = 2\pi/\tau$
($\tau = 365$~d), and $t$ is measured from June 2$^{\rm nd}$.  From
the differential rates evaluated at June, $S_J \equiv S(t=0)$, and
December, $S_D \equiv S(t=\tau/2)$ expressions for $S_{m}$ and $S_{\rm
o}$ can be obtained,
\begin{equation}
 S_m \;=\; \frac{1}{2}\, \left[ S_J \,-\, S_D \right]
  \hspace{0.7cm} S_{\rm o} \;=\; \frac{1}{2}\, \left[ S_J \,+\, 
  S_D \right].
\end{equation}
Assuming that $S_m \ll S_{\rm o}$, the theoretical
signal-to-noise ratio is defined to be,

\begin{equation} (s/n)_{th} \;\equiv \; \frac{S_m}{\sqrt{S_{\rm o}}}
\, \sqrt{MT} \hspace{0.7cm} (S_m \ll S_{\rm o}), \label{snt}
\end{equation} where the product $MT$ is the exposure of the detector,
namely its mass times the overall exposition time to the WIMP flux.
This is the function we need to maximize.  Notice that in (\ref{snt})
it is assumed that no spurious background is present. A more
realistic case will be considered in the next section.

The limits of integration of equation (\ref{st}) are arbitrary and
therefore the theoretical signal-to-noise ratio (\ref{snt}) is
effectively energy dependent,
\begin{equation}
  (s/n)_{th}(E) \; = \; \frac{S_m(E)}{\sqrt{S_{\rm o}(E)}} \,
  \sqrt{MT} . \label{sntE}
\end{equation}

In the case of germanium detectors the crossing energies, $E_c$,
usually
lie between the energy threshold of the detector, $E_i$, and an upper
limit, $E_f$\footnote{An upper limit of 50~keV seems adequate since
the energy deposited in a germanium detector by a 10~TeV WIMP with a
mean velocity of 340~km~s$^{-1}$ is of that order.}, representative of
the highest energy depositions produced by WIMP's.  The maximum
signal-to-noise ratio, for low-mass WIMP's, are obtained when the
differential rates of June and December are integrated between $E >
E_c$ and $E_f$, since the spectra there differ over a larger fraction
of the total energy window, $(E_i, E_f)$.  For heavier WIMP's,
however, with larger values of $E_c$, the maximum signal-to-noise
ratios will be attained when the differential rates is integrated in
the low energy region, from $E_i$ to $E < E_c$.

The theoretical signal-to-noise ratio for 1~kg~yr exposure as a
function of the WIMP mass is shown in Fig. 2; only integrals from
$E$ to $E_f$ are plotted.  For each integration interval the
corresponding ratio is calculated and plotted at $E$, while $E$ is
varied between $0$ and $E_f$.  From the figure one can extract, for
each WIMP mass, the value of $E$ that maximizes the signal-to-noise
ratio, $E_{s/n}$.

\vspace{-1.cm}

\begin{figure}
\centering
\epsfxsize=9.0truecm
\epsfysize=6.8truecm
\epsffile{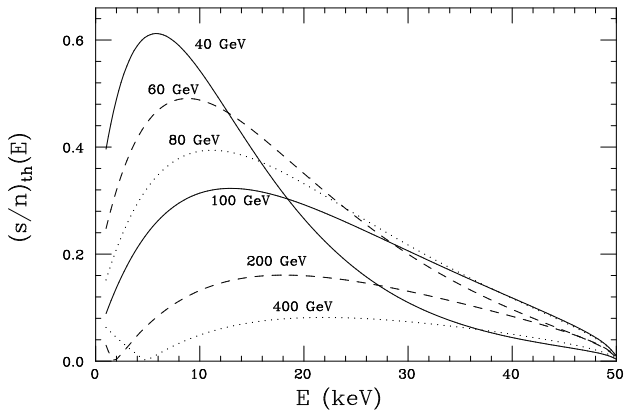}
\caption{Theoretical signal-to-noise ratios in Ge for 1~kg~yr for several
WIMP masses. The ratios were calculated for all energy intervals
running from $E$ to $E_f = 50$~keV and plotted at $E$.\label{fig2}}
\end{figure}

In a similar fashion, we define $E'_{s/n}$ as the value of $E$ that
maximizes $(s/n)_{th}$ when $E$ varies from $E_i$ up to $E$.  In
Fig.~3 the values of $E_{s/n}$ and $E'_{s/n}$ are plotted as a function
of the WIMP mass.

\vspace{-1.cm}

\begin{figure}
\centering
\epsfxsize=9.0truecm
\epsfysize=6.8truecm
\epsffile{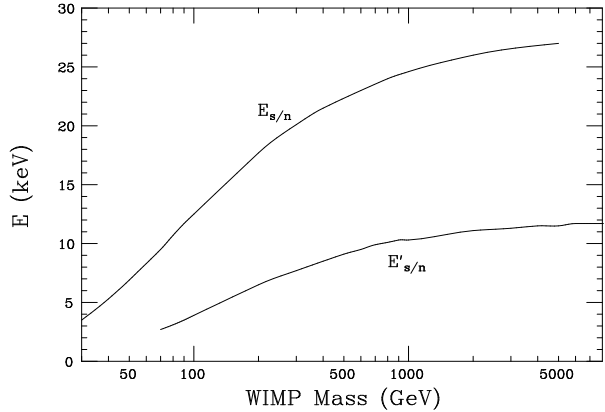}
\caption{Values of $E_{s/n}$ ($E'_{s/n}$) in Ge maximizing the theoretical
signal-to-noise ratios when the differential rate is integrated between
$E$ and $E_f$ ($E_i$ and $E$).\label{fig3}}
\end{figure}

For the next section a quantitative measure of the annual effect is
needed.  We calculate therefore, the percentage ratio of the modulated
over the unmodulated signal, $\alpha = S_m/S_{\rm o}$, and plot it as
a function of the mass of the WIMP in Fig. 4. The solid-line is
obtained using the energy intervals that maximize $(s/n)_{th}(E)$ with
relative intensities between 2~--~8\% (solid line).  In the same
figure a similar ratio but integrated between an arbitrary minimum
energy, $E_i = 2.5$~keV, and $E_f = 50$~keV (dashed line) is also
shown to illustrate the importance of the choice of the energy
interval.  This second curve shows a dramatic decrease in the
modulation percentage for the mass that symmetrizes, with respect to
$E_c$, the interval

\vspace{-1.cm}

\begin{figure}
\centering
\epsfxsize=9.0truecm
\epsfysize=6.8truecm
\epsffile{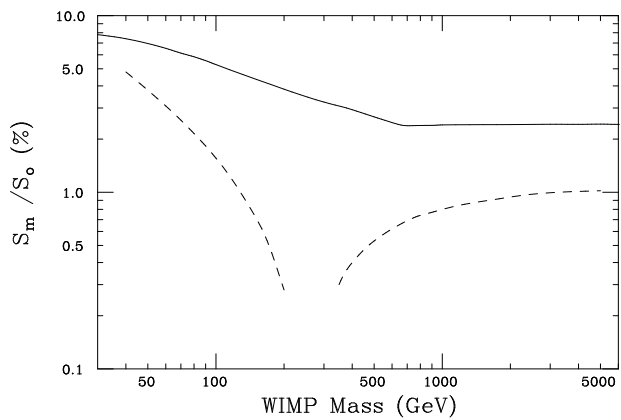}
\caption{Percentage of predicted annual modulation in Ge in two illustrative
cases. An optimized case (solid line) using the energy intervals that
maximize $(s/n)_{th}(E)$ and a non-optimized one (dash line) where $E_i$
has been arbitrarily selected to be 2.5~keV.\label{fig4}}
\end{figure}

\noindent
of integration ($\approx 300$~GeV).  Table
\ref{table2} lists the values of $E_{s/n}$ and $E'_{s/n}$, the second
with three different threshold energies, and the corresponding values
of $S_m/S_{\rm o}$.

\section{Detection sensitivity}

Generalizing the concepts introduced in the previous section the total
rate under experimental conditions can be expressed as,
\begin{equation}
S_{tot} = B + S_{\rm o} + S_m \cos(\omega t) ,
\end{equation}
where $B$ is the rate of spurious background.  The {\em experimental}
signal-to-noise ratio ($S_m \ll S_{\rm o}$) is then,

\begin{equation}
(s/n)_{exp} \; =\; \frac{S_m}{\sqrt{S_{tot}}}\, \sqrt{MT} \; \approx \;
       \frac{\alpha S_{\rm o}}{\sqrt{B + S_{\rm o}}} \, \sqrt{MT},
       \label{sn}
\end{equation}
where $\alpha = S_m/S_{\rm o}$ is as in Table \ref{table2}.  Relation
(\ref{sn}) is always valid; however, determining the real magnitude 

\begin{table}
\caption{Energies that maximize the signal-to-noise ratios in Ge. 
Depending on the WIMP mass the signal-to-noise ratio was obtained integrating
the differential rates of June and December, $dR$, in two cases:  when
$\int_{E}^{E_f} dR$ was considered, $E_{s/n}$ coresponds to the value of
$E$ maximizing the expected signal-to-noise ratio, and when $\int_{E_i}^{E}
dR$ was computed, $E'_{s/n}$  coresponds to the value of $E$ maximizing the
signal-to-noise. $E_f = 50$~keV and $E_i = 0$,~2, and 3~keV were used.}
\vspace{.2cm}

\begin{tabular}{c|l|ccc|c}
$m_{\chi}$ & $E_{s/n}$ & \multicolumn{3}{c|}{$E'_{s/n}$~~(keV)} &
$\alpha \equiv S_m/S_{\rm o}$ \\
(GeV)   &  (keV) & $E_i = 0$ kev & 2 keV & 3 keV & (\%) \\
\hline \hline
   30 &  3.5 &      &      &      & 7.80 \\
   40 &  5.3 &      &      &      & 7.40 \\
   50 &  6.9 &      &      &      & 6.95 \\
   60 &  8.3 &      &      &      & 6.53 \\
   70 &  9.5 &      &      &      & 6.14 \\
   80 & 10.7 &      &      &      & 5.85 \\
   90 & 11.7 &      &      &      & 5.56 \\
  100 & 12.5 &      &      &      & 5.28 \\
  200 & 17.7 &  6.5 &      &      & 3.83 \\
  300 & 20.1 &  7.7 &      &      & 3.24  \\
  400 & 21.5 &  8.5 &  9.4 &      & 2.93  \\
  500 &      &  9.1 & 10.0 & 10.4 & 2.67  \\
  600 &      &  9.5 & 10.4 & 10.8 & 2.48  \\
  700 &      &  9.9 & 10.6 & 11.2 & 2.38  \\
  800 &      & 10.1 & 11.0 & 11.4 & 2.39  \\
  900 &      & 10.3 & 11.0 & 11.6 & 2.39  \\
 1000 &      & 10.3 & 11.2 & 11.6 & 2.41  \\
 2000 &      & 11.1 & 12.0 & 12.4 & 2.42  \\
 3000 &      & 11.3 & 12.2 & 12.6 & 2.43  \\
 4000 &      & 11.5 & 12.4 & 12.8 & 2.42  \\
 5000 &      & 11.5 & 12.4 & 13.0 & 2.43  \\
 6000 &      & 11.7 & 12.6 & 13.0 & 2.42  \\
 7000 &      & 11.7 & 12.6 & 13.0 & 2.43  \\
 8000 &      & 11.7 & 12.6 & 13.0 & 2.43  \\
 9000 &      & 11.7 & 12.6 & 13.0 & 2.43  \\
\end{tabular}
\label{table2}
\end{table}

\noindent
of the background implies modeling very accurately several possible
sources of spurious pulses and their intensities (which is rather
difficult considering the ambiguities involved).  At this point, one
generally assumes that $S_m$ is simply a given fraction, $\alpha '$,
of the {\em total} rate $B + S_{\rm o}$ and not just a fraction of
$S_{\rm o} $; if this is done then $s/n \approx \alpha' \sqrt{(B +
S_{\rm o})MT}$.  Notice, however, that an experiment with a large
background could drive $s/n$ to artificially large values leading to
a wrong estimate of the sensitivity of the experiment.

To solve this apparent inconsistency we need a way to distinguish
between $B$ and $S_{\rm o}$.  We propose to use the experiment that
provides the best exclusion plot as an upper bound to the unmodulated
dark matter signal, $S_{\rm o}$.  In other words, we consider the rate
of such experiment as free of background ($B=0$).  We call this rate
$S_L$ and, consequently, for this experiment $S_{\rm o} = 
S_L$\footnote{Of course, we do not know for sure if $S_L$ is purely a
dark-matter signal or still has some spurious background hidden in it.
The best we can do is to assume that it is pure dark matter and
estimate the exposure required to detect a modulated signal of
value $\alpha$.  If no modulation is found, it is either because there
is none, or because $S_L \neq S_{\rm o}$, and consequently our
original assumption of a negligible background was wrong.  The recipe
then calls for a further reduction of the background and an increase
in the exposure.}. An arbitrary experiment with a total rate
higher than that of the best exclusion, will then have $S_{tot} = B
+ S_L$, where $B$ now represents the difference in total rates between
the two.  The necessary exposure for this arbitrary experiment, to
achieve a level of sensitivity $\alpha$, can be estimated from
(\ref{sn}).  Thus,

\begin{equation}
  MT \;=\; \left( \frac{s/n}{\alpha} \right)^2 \, \frac{1}{S_L}\,
            \left(1\, +\, \frac{B}{S_L} \right) .
   \label{mt2}
\end{equation}
The best scenario, just mentioned, is that where $\gamma \equiv
S_{tot}/S_L \approx 1$, $B \approx 0$, and the minimum required
exposure is therefore,

\begin{equation}
  MT_{min} \;=\; \left( \frac{s/n}{\alpha} \right)^2 \, \frac{1}{S_L} .
   \label{mtb}
\end{equation}
Conversely, when $S_{tot}/S_L \gg 1$,
longer exposures are required to achieve a similar sensitivity. In
this case the necessary exposure is,
\begin{equation}
   MT \;=\;  \gamma \, MT_{min} \; \approx\; \frac{B}{S_L} \, MT_{min}
   \hspace{1.cm} (\gamma \gg 1). \label{mta}
\end{equation}

To obtain $S_L$ we resort to the exclusion plot, $\sigma(m_{\chi})$,
using the relation,

\begin{equation}
 S_L \approx \int_{E_{s/n}}^{E_f}\, \frac{dR[\sigma(m_{\chi})]}{dE}\, dE .
     \label{cond}
\end{equation}
The
differential rate, $dR(\sigma)/dE$, on the left-hand side is calculated
as in section II but using the experimental exclusion, $\sigma(m_\chi)$,
instead of the theoretical zero-momentum cross section,

\begin{equation}
  \frac{d\sigma}{dT} \;=\; \frac{\sigma(m_{\chi})}{T_{max}}\, F^2(k).
\end{equation}
To maintain self-consistency we integrate the differential rate using
the energy limits and the value of $\alpha$ derived in the previous
section for each WIMP mass.  Thus, the predicted signal is
renormalized to make it compatible with the values of
$\sigma(m_{\chi})$ obtained from the exclusion plot.

Solving from equations (\ref{mtb}) and (\ref{cond}), the minimum
exposure can be expressed as

\begin{equation}
  MT_{min} \;=\; \left( \frac{s/n}{\alpha} \right)^2\,
       \left( \int_{E_{s/n}}^{E_f}\, \frac{dR[\sigma(m_{\chi})]}{dE}\, dE
       \right)^{-1}. \label{mt}
\end{equation}

\vspace{-1.cm}

\begin{figure}
\centering
\epsfxsize=9.0truecm
\epsfysize=6.8truecm
\epsffile{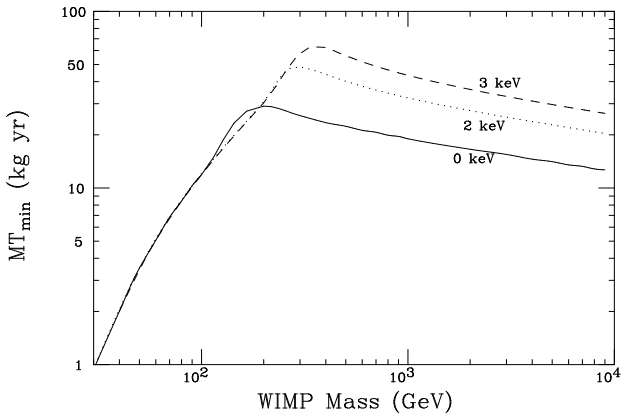}
\caption{Predicted minimum exposures in Ge as a function of WIMP mass for 
three energy thresholds, $E_i = 0$, 2, and 3~keV. The curves were derived 
using the exclusion plot of Ref.~\protect\cite{Bernabei}.\label{fig5}}
\end{figure}

\noindent
In Fig. 5 and Table \ref{table3}, we show the minimum exposures (in
kg~yr) for the lowest rate experiment as a function of WIMP mass based
on the modulation percentages calculated in Table \ref{table2}.  As a
reasonable criterion to distinguish the signal, we have used that $s$
has to be at least $2\sigma$ larger than the statistical uncertainty
or, what is the same, that $s/n = 2$.  The figure and table were
obtained using for $\sigma(m_{\chi})$ the data from Ref.
\cite{Bernabei}, properly rescaled to germanium nuclei. As mentioned in
the previous section, for low-mass WIMP's, the energy intervals maximizing
the signal-to-noise ratios tend to be located {\em above} the crossing
energies. For these masses the minimum exposures are obtained integrating
relation (\ref{mt}) between $E_{s/n}$ and $E_f$. Heavier WIMP's maximize
their $s/n$ at energy intervals {\em below} their crossings. Therefore,
the three curves corresponding to different energy thresholds were obtained
integrating (\ref{mt}) between $E_i = 0$,~2, and 3~keV and $E'_{s/n}$.
Notice that the sensitivity decreases roughly as $\alpha^{-2}$.

For an arbitrary experiment with a total rate $S_{tot}$ the required
exposure can be estimated from equation (\ref{mta}), in units of the
exposure of the lowest rate experiment. The difference between the two
can be attributed to a spurious background of magnitude $B
= (\gamma - 1) S_L$.

An analogous calculation can be done for a NaI detector using the
measured relative efficiencies (quenching factors, $q(I) = 0.09$, $q(Na) = 
0.30$ \cite{Lewin}) and generalazing the formulae of section II to the case 
of a two-nuclei target. Similar results to those of previous sections are found
with the significant distinction that because of the rather low quenching 
factor of iodine, the values of $E_c$, $E_{s/n}$, and $E'_{s/n}$ occur at low 
energies ($< 10$~keV). For NaI a lower value of $E_f$ was used (30~keV). 
Furthermore, since the iodine nucleus cross section is larger than that of 
sodium, iodine is the main contributor to the total rate at low energies
(at high energies its nuclear form factor strongly diminishes the
interaction rate).

The minimum exposure for NaI as a function of WIMP mass is shown in Fig.~6. 
Due to the large cross section of iodine smaller exposures than in the case of
germanium are expected. Nevertheles, as can be appreciated from the figure,
$MT_{min}$ is now more dependent on the threshold energy of the detector.

\vspace{-1.cm}

\begin{figure}
\centering
\epsfxsize=9.0truecm
\epsfysize=6.8truecm
\epsffile{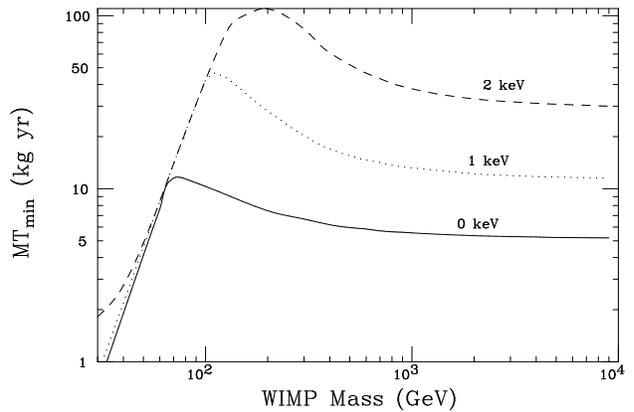}
\caption{Predicted minimum exposures in NaI as a function of WIMP mass for 
three energy thresholds, $E_i = 0$, 1, and 2~keV according to the exclusion 
plot of Ref.~\protect\cite{Bernabei}.\label{fig6}}
\end{figure}

\section{Conclusions}

We have performed a quantitative prediction of annual modulation rates
for Ge and NaI detectors.  To this aim, we found first the adequate
energy intervals where to integrate the rate so that the theoretical
signal-to-noise ratio is at its maximum.  Using these optimized
theoretical rates and assuming that the lowest experimental rate
achieved so far contains no spurious events we then estimated the
minimum exposure required for the predicted modulated rates as a 
function of WIMP mass.  From the results obtained we can draw a few
conclusions.  To avoid artificially large signal-to-noise ratios, an
arbitrary experiment with a total rate, $S_{tot}$, has to be compared
with the best experiment, $S_L$, and should attribute the difference
in counting rate to the presence of a background of the order of $B =
(\gamma - 1) S_L$,~$\gamma = S_{tot}/S_L$.  This arbitrary experiment
requires an exposure of $MT = \gamma MT_{min}$ to reach the same level
of sensitivity as the best one.  Current exclusion plots correspond to
exposures in the range $MT \sim 10 - 60$~kg~yr for masses $m_{\chi} >
100$~GeV (Ge) and between $MT \sim 5 - 90$~kg~yr for masses  $m_{\chi} >
300$~GeV (NaI). A strong dependence in NaI of the minimum exposure  
with threshold energy is also predicted. In Ge, the sensitivity 
is independent from the threshold energy for low WIMP masses.

A recent claim by the DAMA collaboration \cite{Rita} deserves a comment. 
The best exclusion plot used throughout this work corresponds to an 
experimental mean rate of at least, $\langle S_L \rangle \approx 2$~c/kg~d
(energy region 2~-~20~keV), obtained 
by the same group after rejecting events by pulse shape analysis. For the 
annual modulation analysis, however, no rejection is allowed and therefore
the corresponding experimental rate is higher, $S_{tot} \approx 
1$~c/keV~kg~d $\times~18$

\widetext

\begin{table}
\caption{Rates normalized to be compatible with the best exclusion plot
of Bernabei {\em et al.}~\protect\cite{Bernabei} and corresponding exposures
for several WIMP masses in Ge for the intervals of integration
of Table II. }
\vspace{.3cm}

\begin{tabular}{c| c| c c c| c |c c c}
$m_{\chi}$ & $\int_{E_{s/n}}^{E_f} dR(\sigma)$ &  \multicolumn{3}{c|}{
$\int_{E_i}^{E'_{s/n}} dR(\sigma)
$~~~(c/kg d)} & $MT_{min}$ & \multicolumn{3}{c}{$MT_{min}$~~~(c/kg y)} \\
(GeV)   &  (c/kg d) & $E_i = 0$~kev & 2 keV & 3 keV & (c/kg y) & $E_i = 0$~kev
& 2 keV & 3 keV \\
\hline \hline
   30 & 2.035 &       &       &       &  0.94  &         &         &         \\
   40 & 1.057 &       &       &       &  2.01  &         &         &         \\
   50 & 0.681 &       &       &       &  3.52  &         &         &         \\
   60 & 0.534 &       &       &       &  5.06  &         &         &         \\
   70 & 0.447 &       &       &       &  6.83  &         &         &         \\
   80 & 0.398 &       &       &       &  8.43  &         &         &         \\
   90 & 0.360 &       &       &       & 10.29  &         &         &         \\
  100 & 0.344 &       &       &       & 11.91  &         &         &         \\
  200 & 0.253 & 0.737 &       &       & 30.75  &  29.07  &         &         \\
  300 & 0.223 & 0.793 &       &       & 48.42  &  25.67  &         &         \\
  400 & 0.208 & 0.853 & 0.648 &       & 62.50  &  23.44  &  43.98  &         \\
  500 &       & 0.888 & 0.688 & 0.594 &        &  22.36  &  40.20  &  56.06  \\
  600 &       & 0.928 & 0.717 & 0.622 &        &  21.21  &  37.68  &  51.84  \\
  700 &       & 0.953 & 0.736 & 0.651 &        &  20.68  &  35.86  &  48.87  \\
  800 &       & 0.984 & 0.767 & 0.667 &        &  19.90  &  34.45  &  46.63  \\
  900 &       & 0.997 & 0.775 & 0.687 &        &  19.58  &  33.32  &  44.86  \\
 1000 &       & 1.016 & 0.795 & 0.695 &        &  18.99  &  32.28  &  43.43  \\
 2000 &       & 1.155 & 0.912 & 0.803 &        &  16.57  &  27.44  &  36.15  \\
 3000 &       & 1.226 & 0.979 & 0.863 &        &  15.49  &  25.19  &  33.00  \\
 4000 &       & 1.306 & 1.036 & 0.914 &        &  14.56  &  23.78  &  31.05  \\
 5000 &       & 1.341 & 1.074 & 0.960 &        &  14.10  &  22.76  &  29.67  \\
 6000 &       & 1.408 & 1.119 & 0.989 &        &  13.51  &  21.97  &  28.62  \\
 7000 &       & 1.432 & 1.148 & 1.015 &        &  13.26  &  21.34  &  27.77  \\
 8000 &       & 1.477 & 1.174 & 1.038 &        &  12.83  &  20.81  &  27.06  \\
 9000 &       & 1.493 & 1.198 & 1.059 &        &  12.67  &  20.35  &  26.46  \\
\end{tabular}
\label{table3}
\end{table}

\narrowtext

\noindent
keV $\approx 18$~c/kg~d. This implies the presence of a spurious
background that increases the necessary exposure required to detect a 
modulation by a factor $\gamma = S_{tot}/S_L \approx 9$. According to 
our calculations then, the required exposure ($\approx 65$~kg~yr for a
60~GeV WIMP) was not achieved by the DAMA group since, at the 
moment of their announcement, they had analyzed a statistics of only 
12.5~kg~yr. It is interesting to note however, that our results predict that 
the DAMA collaboration with its 115.5~kg detector should be sensitive to low ($ \lesssim 90$~GeV)
as well as high-mass ($\gtrsim 1$~TeV) WIMPS in approximately three years of
data acquisition.
 
Future improvements in background reduction will certainly push
further down the limits of dark matter signal and at the same time
will make the small annual fluctuations harder to distinguish.
Nevertheless, in view of Figs. 5, 6, and equation (\ref{mta}),
higher-background experiments with large masses will still be suitable 
to explore the low and high mass regions.

\newpage

\acknowledgements{The author would like to express his gratitude
to A.~O.~Gattone and the rest of the TANDAR group: D.
Abriola, D.~E.~Di~Gregorio, C.~K. Gu\'erard, and H. Huck for helpful
discussion and comments. \\
\\
$^{\rm *}$Pressent address: Laboratory for High
Energy Physics, University of Bern, Sidlerstrasse 5, CH 3012, Bern, 
Switzerland}

\end{document}